\pdfoutput=1
\documentclass[twocolumn,prd,superscriptaddress,preprintnumbers,nofootinbib]{revtex4}

\usepackage{graphicx}
\usepackage{epsfig}
\usepackage{bm}

\usepackage{latexsym,amssymb,amsmath,amsfonts,amssymb,txfonts,pxfonts,wasysym,float}
\usepackage{color}

\usepackage{enumitem}

\newcommand{\postscript}[2]{\setlength{\epsfxsize}{#2\hsize}
   \centerline{\epsfbox{#1}}}

\usepackage[usenames,dvipsnames]{xcolor}
\definecolor{orange}{cmyk}{0,0.5,1,0}
\definecolor{rossoCP3}{cmyk}{0,.88,.77,.40}
\definecolor{graa}{rgb}{0.8,0.8,0.8}
\definecolor{blaa}{rgb}{0.2,0.2,0.6}

\begin{document}

\title{\color{rossoCP3} Prospects for macroscopic dark matter detection at space-based and suborbital experiments}

\author{Luis A. Anchordoqui}

\affiliation{Department of Physics and Astronomy,  Lehman College, City University of
  New York, NY 10468, USA
}

\affiliation{Department of Physics,
 Graduate Center, City University
  of New York,  NY 10016, USA
}

\affiliation{Department of Astrophysics,
 American Museum of Natural History, NY
 10024, USA
}

\author{Mario E. Bertaina}
\affiliation{Dipartimento di Fisica, Universit\'a di Torino, Torino 10125, Italy}

\author{Marco Casolino}
\affiliation{Istituto Nazionale di Fisica Nucleare, Section of Roma Tor Vergata, Italy}

\author{Johannes Eser}
\affiliation{Department of Astronomy \& Astrophysics, KICP, EFI,  University of Chicago, Chicago, IL 60637, USA}

\author{John~\nolinebreak~F.~\nolinebreak~Krizmanic}
\affiliation{Astrophysics Science Division, NASA Goddard Space Flight Center, Greenbelt, MD 20771, USA}
\affiliation{University of Maryland, Baltimore County, Baltimore, MD 21250, USA}

\author{Angela~\nolinebreak~V.~\nolinebreak~Olinto}
\affiliation{Department of Astronomy \& Astrophysics, KICP, EFI,  University of Chicago, Chicago, IL 60637, USA}

\author{A. Nepomuk Otte}
\affiliation{Georgia Institute of Technology 837 State Street NW, Atlanta, Georgia 30332-0430, USA}

\author{Thomas C. Paul}
\affiliation{Department of Physics and Astronomy,  Lehman College, City University of
  New York, NY 10468, USA
}

\author{Lech~\nolinebreak~W.~\nolinebreak~Piotrowski}
\affiliation{Faculty of Physics, University of Warsaw, Warsaw, Poland}

\author{Mary Hall Reno}

\affiliation{Department of Physics and Astronomy, University of Iowa, Iowa City, IA 52242, USA}

\author{Fred~\nolinebreak~Sarazin}
\affiliation{Department of Physics,
Colorado School of Mines, Golden, CO 80401, USA}

\author{Kenji~\nolinebreak~Shinozaki}
\affiliation{National Centre for Nuclear Research, Lodz, 90-559, Poland}

\author{Jorge F. Soriano}

\affiliation{Department of Physics and Astronomy,  Lehman College, City University of
  New York, NY 10468, USA
}

\affiliation{Department of Physics,
 Graduate Center, City University
  of New York,  NY 10016, USA
}

\author{Tonia M. Venters}
\affiliation{Astrophysics Science Division, NASA Goddard Space Flight
  Center, Greenbelt, MD 20771, USA}

\author{Lawrence Wiencke}
\affiliation{Department of Physics,
Colorado School of Mines, Golden, CO 80401, USA}

\date{April 2021} 

\begin{abstract}
  \noindent We compare two different formalisms for modeling the
  energy deposition of macroscopically sized/massive quark nuggets
  (a.k.a. macros) in the Earth's atmosphere. We show that for a
  reference mass of 1~g, there is a discrepancy in the macro
  luminosity of about 14 orders of magnitude between the predictions of the
  two formalisms. Armed with our finding we estimate the sensitivity
  for macro detection at space-based (Mini-EUSO and POEMMA) and
  suborbital (EUSO-SPB2) experiments.
\end{abstract}
\maketitle

The conventional textbook dark matter (DM) particle species is assumed to
interact with Standard Model (SM) fields only
gravitationally~\cite{Feng:2010gw}. Actually, the cross
section of the canonical weakly-interacting massive particle (WIMP)~\cite{Steigman:1984ac}
 to scatter from baryons is non-zero though  small enough to be
considered effectively zero for mass scales above a solar
mass~\cite{Dvorkin:2013cea}. Yet, since the WIMP parameter space
keeps shrinking due to null results
at the LHC~\cite{Penning:2017tmb,Rappoccio:2018qxp,Buchmueller:2017qhf} and 
unsatisfactory
answers from the WIMP search program using direct and indirect detection
methods~\cite{Undagoitia:2015gya,Gaskins:2016cha}, the case for
alternative (and especially SM) candidates featuring stronger DM-baryon interactions has grown
stronger, and attracted increasing attention.

Macroscopic DM is a general class of models with DM in a compact and
composite state with a large radius and mass. Nuclearites and its dark
quark nuggets cousins provide two compelling examples. Nuclearites are
macroscopically sized nuggets of strange quark matter which could have
been produced during the QCD phase transition in the early
universe~\cite{Witten:1984rs,Farhi:1984qu,DeRujula:1984axn,Alcock:1988re}. If this were the case then
DM
would have nuclear density,
$\rho_s \sim 3.6 \times 10^{14}~{\rm g/cm^3}$~\cite{Chin:1979yb}. However, this
constraint may be relaxed for the case of dark nuclearites as the dark
quark nugget's energy density may span several orders of magnitude
depending on the confinement scale and the magnitude of the dark
baryon asymmetry~\cite{Bai:2018dxf}. Herein we 
refer to all such macroscopic DM candidates generically as
macros~\cite{Jacobs:2014yca}, and following~\cite{Sidhu:2018auv}, 
we let the macro's energy density to vary in a generous range
$10^6 < \rho_m/{\rm (g/cm^3)} < 10^{15}$.

Elastic scattering allows macros and baryons to exchange
momentum. The process has two undetermined parameters: the mass of the
macro 
$M$ and the interaction cross section
$\sigma$, generally taken to be the geometric cross-sectional area of the
macro. Before proceeding, we pause to note that there remains a large range of the $M - \sigma$ parameter space which is still unprobed by experiment.

If a macro were to traverse through the Earth's atmosphere its energy
deposition would excite the nitrogen molecules of air producing
observable signals at fluorescence detectors. In this Letter we
reexamine the methodology for estimating the
sensitivity for macro detection at space-based and suborbital
experiments. More concretely, we compare one approach for estimating
the macro luminosity originally developed in the
eighties~\cite{DeRujula:1984axn} to a more recent examination of the problem~\cite{Sidhu:2018auv}. We adopt three projects of
the Joint Experiment Missions for Extreme
Universe Space Observatory (JEM-EUSO) as reference in our discussion:
\begin{itemize}[noitemsep,topsep=0pt]
\item the Mini-EUSO detector,
currently taking data on board the International Space Station~\cite{Bacholle:2020emk};
\item the second generation Super-Pressure Balloon long duration flight (EUSO-SPB2), 
which has been approved by NASA to be launched in
2022~\cite{Wiencke:2019vke};
\item the future Probe Of Extreme Multi-Messenger Astrophysics (POEMMA) mission~\cite{Olinto:2020oky}.
\end{itemize}

Like meteoroids, macros are susceptible to rapid heat loss upon
entering the Earth's atmosphere as a result of elastic collisions with
the air molecules. Actually, it is at lower altitudes where the macro
encounters the exponentially increasing atmospheric density and undergoes rapid heating along its path, which expands and
radiates. The power dissipation rate of macros going through the
atmosphere is given by
\begin{equation}
  \frac{dE}{dt} \sim \rho_\mathrm{atm} \ \sigma \ v^3 \,,
\end{equation}
where $\rho_\mathrm{atm}$ is the atmospheric density and $v \sim
250~{\rm km/s}$ is the characteristic velocity of the Sun's galactic rotation~\cite{DeRujula:1984axn}. To describe the atmospheric density variation
we adopt an isothermal atmosphere,
\begin{equation}
  \rho_\mathrm{atm}=\rho_{\mathrm{atm},0} \ \exp\left({-\frac{z}{z^*}}\right) \,,\label{eq:atm}\end{equation}
where $\rho_{\mathrm{atm},0}=10^{-3}~\mathrm{g/cm^3}$ and
$z^*=8~\mathrm{km}$~\cite{Anchordoqui:2018qom}. Now, the power dissipated to
\emph{useful} light is given by
\begin{equation}
  L =\eta \rho_\mathrm{atm}\sigma v^3,
\label{luminosidad}
\end{equation}
where $\eta$ is the luminous efficiency. 

In the model of~\cite{DeRujula:1984axn} it is assumed that the
expanding hot cylinder emits black-body radiation, and its
luminous efficiency is estimated to be
\begin{eqnarray}
\eta_1 & \sim & 2\times10^{-5}\left(\frac{\overline 
    w}{18}\right)^{3/2}\frac{\rho_\mathrm{water}}{\rho_\mathrm{atm}} \sim  0.04 \exp \left(\frac{z}{z^*} \right),
\label{eta1}
\end{eqnarray}
where $\overline w\sim29$ is the average molecular weight of air
molecules and $\rho_\mathrm{water}$ is the water density. Substituting (\ref{eta1}) into (\ref{luminosidad}), the
macro luminosity (for {\it model 1}) can be recast as
\begin{equation}
L_1\approx 15\left(\frac{M}{\mathrm g}\right)^{2/3}x^{-2/3}\,\mathrm W,
\end{equation}
or, as written in~\cite{DeRujula:1984axn}  assuming $x=1$,
\begin{equation}
L_1\approx1.5\times10^{-3}\left(\frac{M}{\mathrm{\mu
      g}}\right)^{2/3}~\mathrm W \,,
\label{L1}
\end{equation}
where
\begin{equation}
  \sigma=2.4\times 10^{-10}\left(\frac{M}{\mathrm
      g}\right)^{2/3}x^{-2/3}~\mathrm{cm^2} \,,
  \label{eq:sigma}
\end{equation}
with  $x\equiv\rho_m/\rho_s$.

An alternative approach to describe the interactions of macros in the atmosphere, which includes a precise determination of 
 the probability for transitions in a nitrogen plasma to produce a
 photon in the $350$ to $400~{\rm nm}$ detection range, has been
 recently 
 developed in~\cite{Sidhu:2018auv}. Within this model the
luminous efficiency is given by
\begin{equation}
\eta_2=2\times 10^5\left(\frac{\sigma}{\mathrm{cm}^2}\right)^2\left(\frac{v}{250\,\mathrm{km/s}}\right)^4\left[\exp\left(-\frac{z}{10\,\mathrm{km}}\right)\right]^4.
\label{eta2-1}
\end{equation}
We note that the exponential comes from the height dependence of
several functions on the atmospheric density, which are modelled as in
(\ref{eq:atm}), but with $z^*=10\,\mathrm{km}$. To remain consistent
with the isothermal atmospheric model adopted for our calculations we write (\ref{eta2-1})  as
\begin{equation}
\eta_2\approx 1.15\times10^{-14}\left(\frac{M}{\mathrm g}\right)^{4/3}x^{-4/3}\exp\left(-\frac{4z}{z^*}\right).
\end{equation}
With this in mind, the macro luminosity for {\it model 2} is given by 
\begin{equation}
  L_2 \approx 4.32\times 10^{-12}\left(\frac{M}{\mathrm g}\right)^2 \
  x^{-2} \ \exp\left(-\frac{5z}{z^*}\right)~\mathrm W.
\label{L2}
\end{equation}
By comparing (\ref{L1}) and (\ref{L2}) it is straightforward to see that {\it for a
  reference mass of 1~g, $x=1$, and $z=z^*$ there is a discrepancy of about 14 orders
of magnitude between the predictions  of the two models}.

\begin{table}
\begin{center}
  \caption{Macro luminosity parameters. \label{tabla}} 
\begin{tabular}{c| c c}
  \hline
  \hline
  ~~~~~~~~~&~~~~~~~~~~~~ {\it model 1} ~~~~~~~~~~~~&~~~~~~~~~~~~ {\it model 2} ~~~~~~~~~~~~\\
\hline
  $\alpha_i$&$2/3$&$2$\\
  $\tilde L_i\,[\mathrm W]$&$15$&$4.32\times10^{-12}$\\
  $f_i$&$1$&$\exp\left(-5 \ z/z^*\right)$\\
  \hline
  \hline
\end{tabular}
\end{center}
\end{table}

\begin{figure*}[tpb]
\begin{minipage}[t]{0.49\textwidth}
\postscript{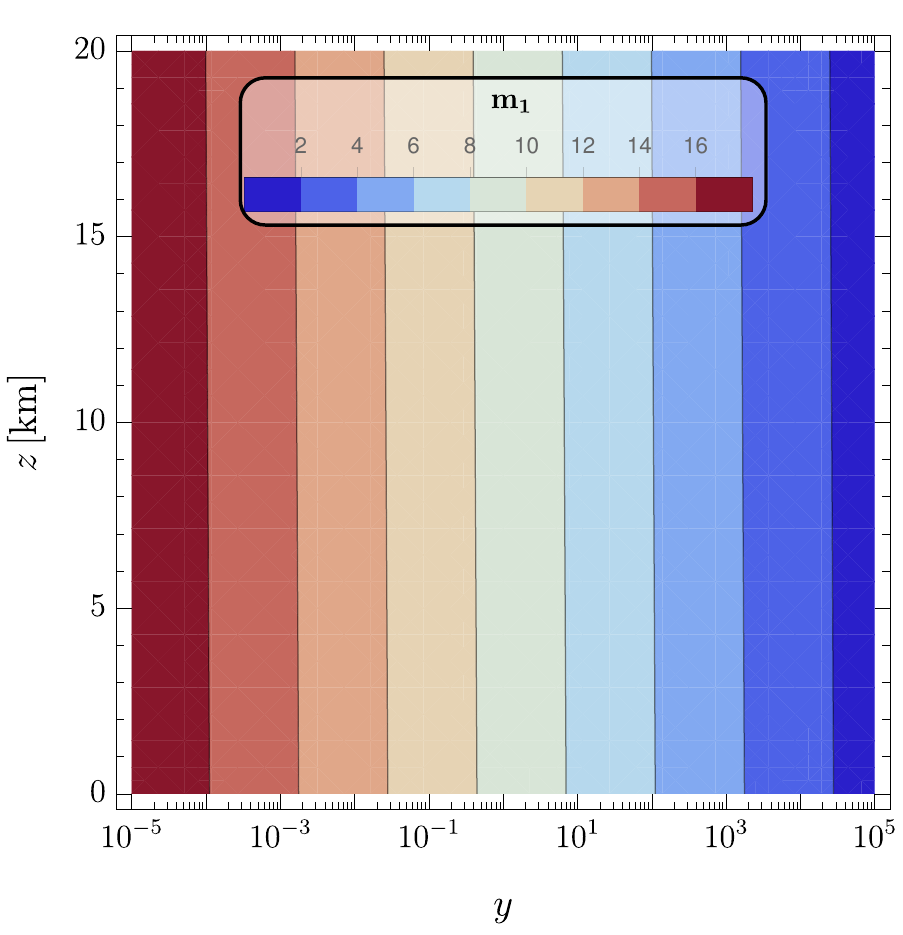}{0.8}
\end{minipage}
\begin{minipage}[t]{0.49\textwidth}
\postscript{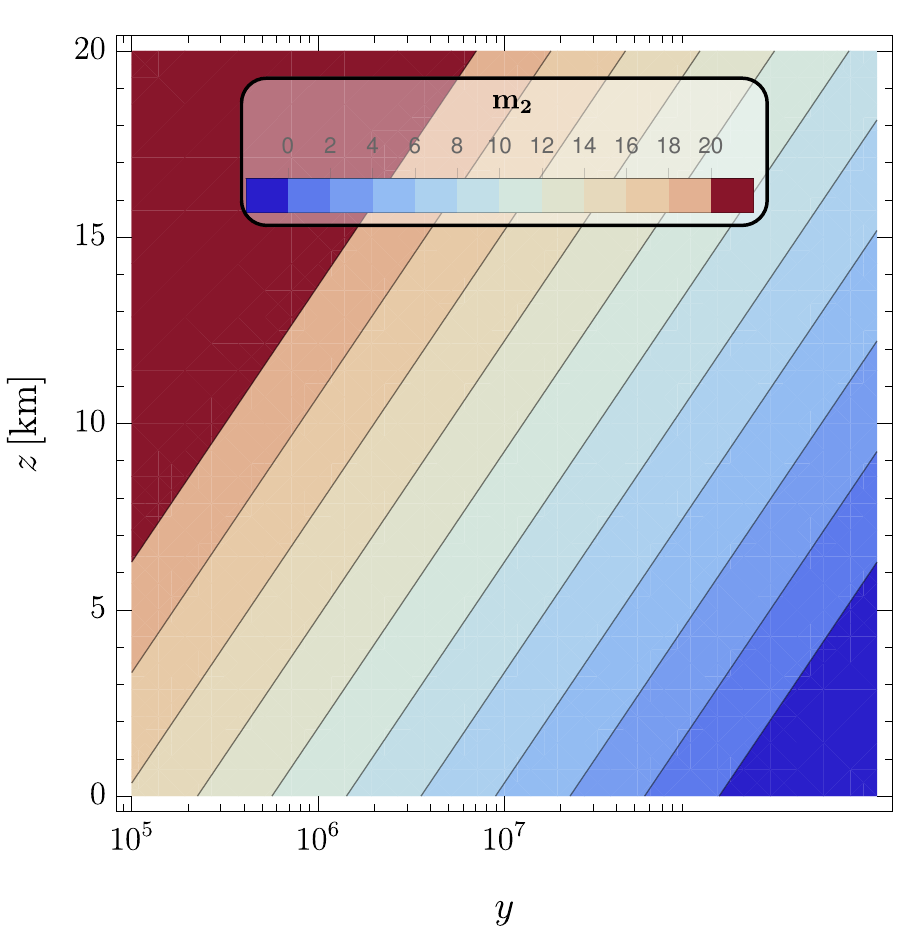}{0.8}
\end{minipage}
\caption{Values of $\mathbf m_1$ (\emph{left}) and $\mathbf m_2$ (\emph{right}) as a function of $y$ and $z$ for POEMMA.}
    \label{fig:poemma-m}
\end{figure*}
\begin{figure*}
\begin{minipage}[t]{0.49\textwidth}
\postscript{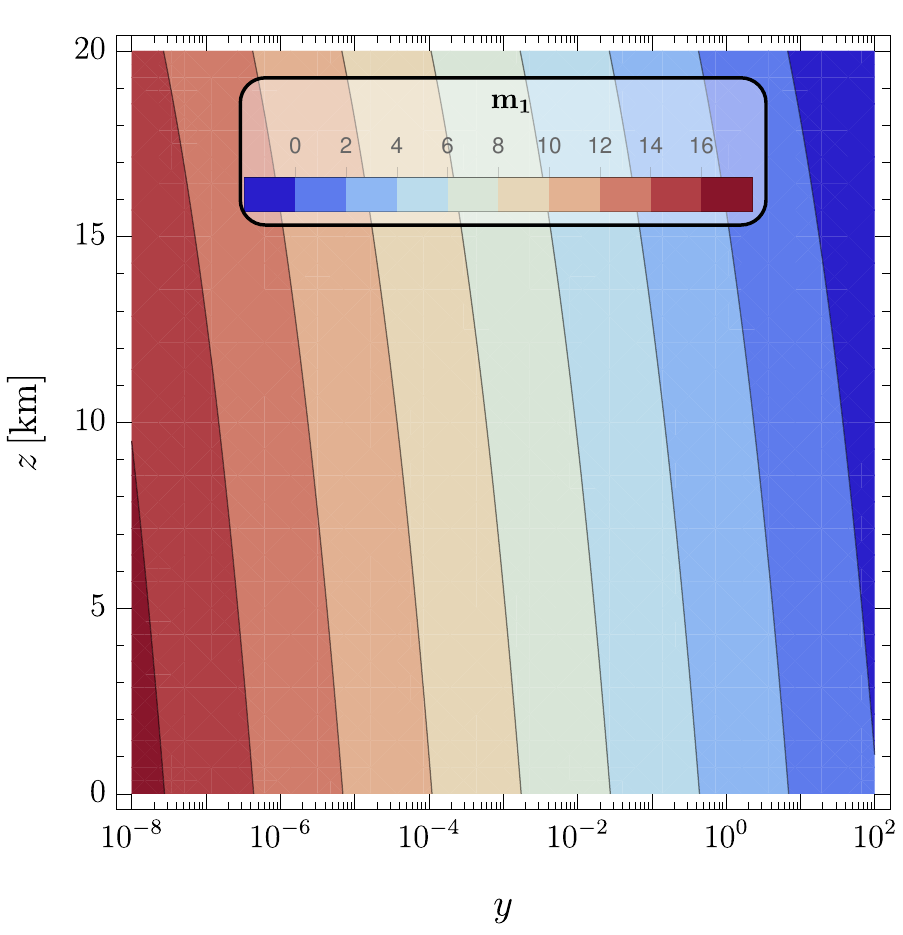}{0.8}
\end{minipage}
\begin{minipage}[t]{0.49\textwidth}
\postscript{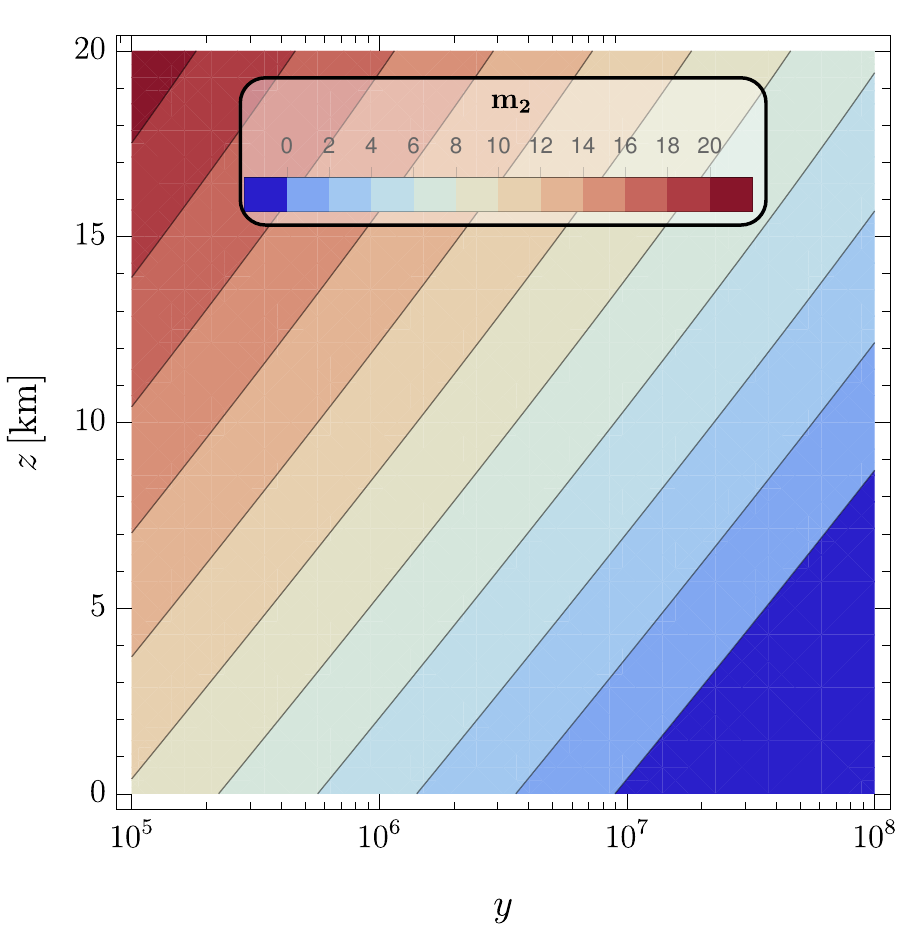}{0.8}
\end{minipage}
\caption{Values of $\mathbf m_1$ (\emph{left}) and $\mathbf m_2$ (\emph{right}) as a function of $y$ and $z$ for EUSO-SPB2.}
    \label{fig:spb-m}
\end{figure*}

The apparent magnitude of an object at a distance $d$ and with
luminosity $L$ is defined as
\begin{equation}
\mathbf{m}=-\frac52\log\frac{L}{4\pi d^2\ell_0},
\label{mW}
\end{equation}
where
$\ell_0\approx2.52\times10^{-8}\,\mathrm{W\,m^{-2}}$~\cite{Weinberg:2008zzc}. For
convenience, we rewrite (\ref{mW}) as
\begin{equation}
  \mathbf m=5\log\frac{d}{d_*}-\frac52\log\frac{L}{4\pi d_*^2\ell_0},
  \label{m}
\end{equation}
where $d_*$ is any reference distance. The luminosity can be rewritten as
\begin{equation}
  L_i=\tilde L_i\left(\frac{M}{\mathrm
      g}\right)^{\alpha_i}x^{-\alpha_i}f_i(z),
\label{Li}
\end{equation}
with parameters given in Table~\ref{tabla}. Substituting (\ref{Li}) into (\ref{m}) we obtain
\begin{eqnarray}\mathbf {m}_i & = &-\frac52\log\frac{\tilde L_i}{4\pi
    d_*^2\ell_0}-\frac{5\alpha_i}{2}\log\left(\frac{M}{\mathrm
                                    g}\frac1x\right)+5\log\frac{d}{d_*} \nonumber \\
  & - & \frac{5}{2} \log f_i(z) \, .
\end{eqnarray}
Following~\cite{DeRujula:1984axn}, we choose a scale 
$d_*=10~\mathrm{km}$ and a vertical observation altitude $h \approx
z+d$, yielding
\begin{subequations}
  \begin{equation}
\mathbf{m}_1=0.811-\frac53\log\frac{M}{\mathrm g}+5\log\frac{h-z}{10\,\mathrm{km}}+\frac53\log x,\label{eq:m1}
\end{equation}
and 
\begin{eqnarray}
\mathbf{m}_2 & = & 32.16-5\log\frac{M}{\mathrm
                   g}+5\log\frac{h-z}{10\,\mathrm{km}}+5\log x
                   \nonumber \\
  & + & \frac{25}{2}\frac{z}{z^*} \frac{1}{\ln10} \ .\label{eq:m2}
  \end{eqnarray}
\end{subequations}
For the purpose of comparison with~\cite{DeRujula:1984axn}, after
setting $x=1$, 
(\ref{eq:m1}) can be recast as
\begin{equation}
\mathbf m_1=10.811-\frac53\log\frac{M}{\mathrm{\mu g}}+5\log\frac{d}{10\,\mathrm{km}}.
\end{equation}
Setting $h = 33~{\rm km}$  it is straightforward to
see by comparing (\ref{eq:m1}) and (\ref{eq:m2}) that for our fiducial
values ($M=1~{\rm g}$, $x =1$, $z=z^*$) the 14 orders of magnitude
discrepancy in luminosity translate into a  macro apparent magnitude
difference $\Delta \mathbf{m} = 36.8$.

Mini-EUSO has demonstrated the capability to detect
meteors~\cite{Bacholle:2020emk}. Indeed Mini-EUSO (at an orbit of
400~km) is sensitive to meteors of
apparent magnitude $\mathbf m = 6$, whereas  POEMMA (at an
orbit of 525~km) will be able to detect meteors
 of $\mathbf m = 10$. These estimates do
not include effects due to potential atmospheric absorption, which
will be discussed elsewhere. Macros travel much faster than meteors
(which being bound to the solar system travel at less than 72~km/s relative to
the Earth) allowing for clean discrimination among the atmospheric
signals. Moreover, clear differences in the meteor/macro light profiles have been observed in numerical simulations~\cite{Adams:2014vgr}.

\begin{figure*}[t]
\begin{minipage}[t]{0.49\textwidth}
\postscript{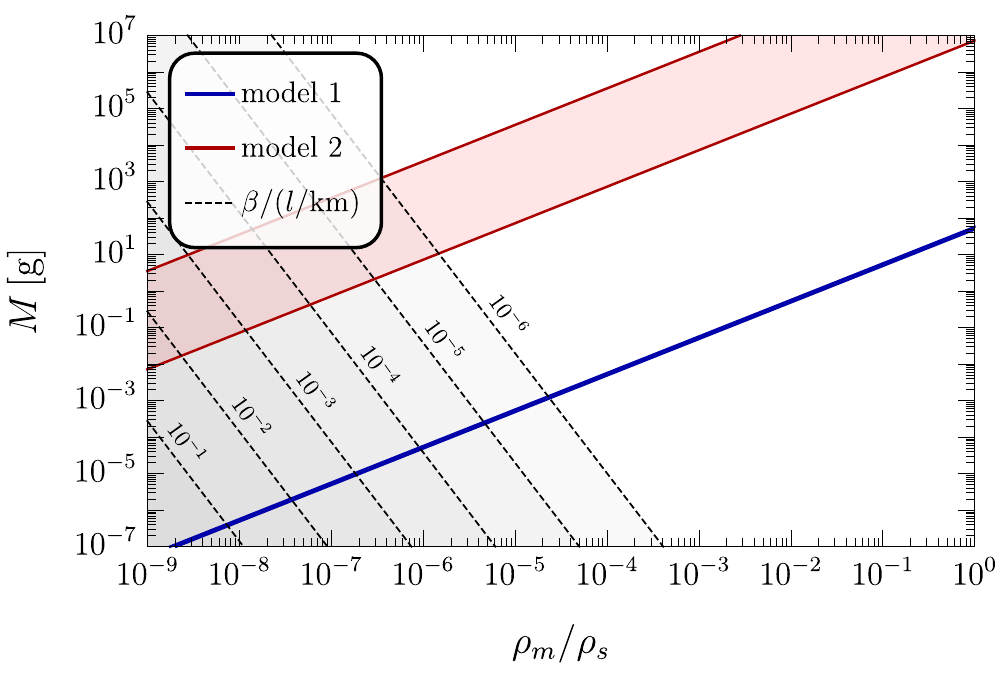}{0.8}
\end{minipage}
\begin{minipage}[t]{0.49\textwidth}
\postscript{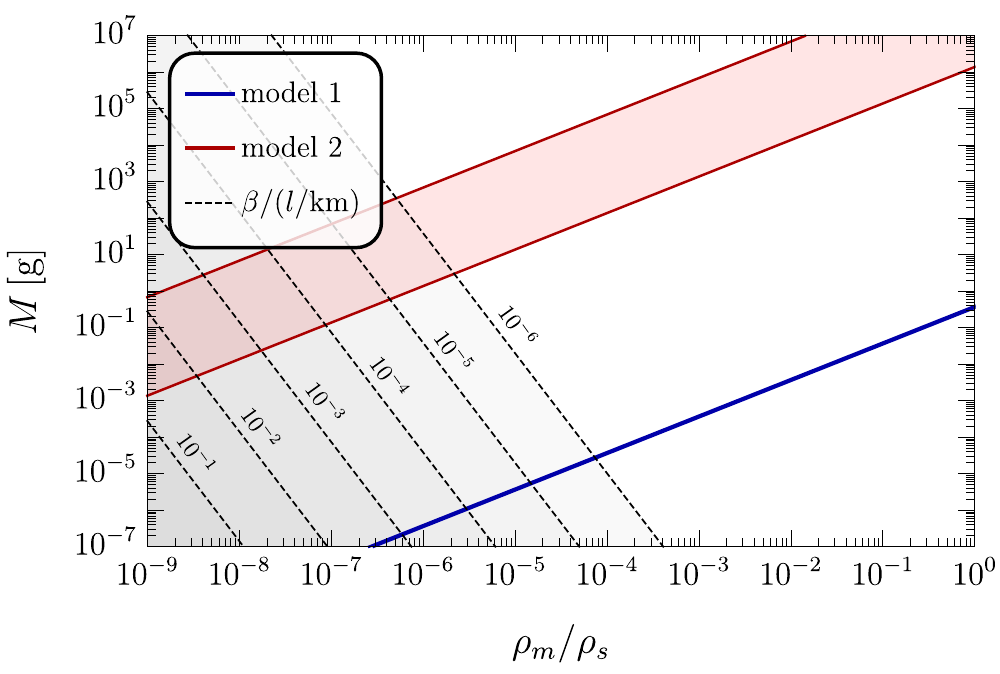}{0.8}
\end{minipage}
\caption{Regions of the $(x,M)$ parameter space for macros at
  altitudes from zero to $20\,\mathrm{km}$ which produce an apparent
  magnitude of $\mathbf m=6$ (left) and $\mathbf m=10$ (right). The regions below the lines produce
  larger magnitudes, so they are harder to observe than those above
  them. The dashed lines show the strength stability constraint
  $\beta\ll1$. }\label{fig:poemma-y}
\end{figure*}

In order to study the observational sensitivity of JEM-EUSO
instruments to $M$ and $x$ under both models we define the parameter
$y\equiv M/(x\,\mathrm g)$. In Fig.~\ref{fig:poemma-m}, we show
constant apparent magnitude contours in the $(y,z)$ plane, considering
the observation altitude of POEMMA spacecraft. For comparison, in
Fig.~\ref{fig:spb-m} we show same contours for EUSO-SPB2, which will
fly at an altitude of about 33~km. There is no appreciable difference
between the contours for Mini-EUSO and POEMMA, but of course they are
sensitive to different apparent magnitudes. An apparent magnitude
$\mathbf{m} = 10$ corresponds to values $y \approx0.37$ for the
first model, and $y\approx5\times10^5$ for the second at $z\approx
z^*$. Substituting this result into (\ref{eq:m2}) we can explore the
sensitivity of Mini-EUSO and POEMMA scanning the $(M,x)$ parameter space. The
results of this exploration are encapsulated in
Fig.~\ref{fig:poemma-y} where we show the corresponding values in the $(x,M)$ parameter space, for both models, in a generous range $z\in[0\,\mathrm{km},20\,\mathrm{km}]$.

The requirement of macro stability as it traverses the
atmosphere sets an additional constraint on the cross section, as $E_b
M/m_b \gg \rho_\mathrm{atm}\sigma v^2 l$, where $l$ is the length
travelled by the macro through the atmosphere, $m_b$ the baryon mass, and the macro binding energy is $E_b\sim10\,\mathrm{eV}[\rho_m/\mathrm{(g/cm^3})]^{3/7}$~\cite{Sidhu:2019fgg}. Substituting $\sigma$ and $\rho_\mathrm{atm}$ from (\ref{eq:atm}) and (\ref{eq:sigma}) this translates into a condition $\beta\ll1$, where
\begin{equation}
    \beta\equiv\frac{\rho_\mathrm{atm} v^2\sigma l}{E_b M/m_b}\approx 9\times10^{-13}\left(\frac{M}{\mathrm g}\right)^{-1/3}x^{-23/21}\frac{l}{\mathrm{km}},
\end{equation}
and where we have considered the upper bound on the density,
$\rho_\mathrm{atm}=\rho_{\mathrm{atm},0}$ to be conservative. The
lines with constant $\beta/l$, which allow to determine the excluded
areas for multiple lengths, are shown in Fig.~\ref{fig:poemma-y}. An upper bound for $l$ may be set by assuming a trajectory tangent to the Earth's surface that starts and ends at a height $z$ over the surface. In such case, $(R_\oplus+z)^2=R_\oplus^2+(l/2)^2$, which yields $l\approx\sqrt{8R_\oplus z}$, with a value of a few hundreds depending on the chosen $z$. A very conservative overestimate, for $z\sim20\,\mathrm{km}$, is $l\sim1000\,\mathrm{km}$.

All in all, we can conclude that:
\begin{itemize}[noitemsep,topsep=0pt]
\item Mini-EUSO is sensitive to macros of  $x \sim 1.3 \times 10^{-8}$
  for $M \agt 1~{\rm g}$, and macros of $x\sim 1$ for $M \agt 8.1 \times 10^7~{\rm g}$;
\item the future POEMMA mission will be
sensitive to macros of  $x \sim 6.1 \times 10^{-8}$ for $M \agt 1~{\rm g}$, and macros of
$x\sim 1$ for $M \agt 1.6\times 10^7~{\rm g}$.
\end{itemize}

\acknowledgments{This work has been partially supported by NASA Grants
  80NSSC18K0464 (L.A.A., T.C.P.,  J.F.S.), 80NSSC18K0246 (J.E.,
  A.V.O.), 80NSSC19K0626 (J.F.K.), 80NSSC18K0477  (F.S., L.W.),
  17-APRA17-0066 (T.M.V.), and DoE Grant DE-SC-0010113 (M.H.R). M.E.B. is
supported by Compagnia di San Paolo within the project 'ex-post-2018.'
K.S. is supported by the National Science Centre in Poland, Grant 
2020/37/B/ST9/01821. Any opinions,
findings, and conclusions or recommendations expressed in this
material are those of the authors and do not necessarily reflect the
views of the NASA or DoE.}


\begin{thebibliography}{99}

\bibitem{Feng:2010gw}
J.~L.~Feng,
 {\color{rossoCP3} Dark matter candidates from Particle Physics and methods of detection},
Ann. Rev. Astron. Astrophys. \textbf{48}, 495-545 (2010)
doi:10.1146/annurev-astro-082708-101659
[arXiv:1003.0904 [astro-ph.CO]].


  
\bibitem{Steigman:1984ac}
  G.~Steigman and M.~S.~Turner,
    {\color{rossoCP3} Cosmological constraints on the properties of weakly interacting massive particles},
  Nucl.\ Phys.\ B {\bf 253}, 375 (1985).
  doi:10.1016/0550-3213(85)90537-1




  
\bibitem{Dvorkin:2013cea}
C.~Dvorkin, K.~Blum and M.~Kamionkowski,
{\color{rossoCP3} Constraining mark matter-baryon scattering with linear cosmology},
Phys. Rev. D \textbf{89}, no.2, 023519 (2014)
doi:10.1103/PhysRevD.89.023519
[arXiv:1311.2937 [astro-ph.CO]].


\bibitem{Penning:2017tmb}
B.~Penning,
 {\color{rossoCP3} The pursuit of dark matter at colliders\textemdash{}an overview},
J. Phys. G \textbf{45}, no.6, 063001 (2018)
doi:10.1088/1361-6471/aabea7
[arXiv:1712.01391 [hep-ex]].



\bibitem{Rappoccio:2018qxp}
S.~Rappoccio,
 {\color{rossoCP3} The experimental status of direct searches for exotic physics beyond the standard model at the Large Hadron Collider},
Rev. Phys. \textbf{4}, 100027 (2019)
doi:10.1016/j.revip.2018.100027
[arXiv:1810.10579 [hep-ex]].


\bibitem{Buchmueller:2017qhf}
  O.~Buchmueller, C.~Doglioni and L.~T.~Wang,
   {\color{rossoCP3} Search for dark matter at colliders},
Nature Phys. \textbf{13}, no.3, 217-223 (2017)
doi:10.1038/nphys4054
[arXiv:1912.12739 [hep-ex]].


\bibitem{Undagoitia:2015gya}
T.~Marrod\'an Undagoitia and L.~Rauch,
 {\color{rossoCP3} Dark matter direct-detection experiments},
J. Phys. G \textbf{43}, no.1, 013001 (2016)
doi:10.1088/0954-3899/43/1/013001
[arXiv:1509.08767 [physics.ins-det]].



\bibitem{Gaskins:2016cha}
J.~M.~Gaskins,
 {\color{rossoCP3} A review of indirect searches for particle dark matter},
Contemp. Phys. \textbf{57}, no.4, 496-525 (2016)
doi:10.1080/00107514.2016.1175160
[arXiv:1604.00014 [astro-ph.HE]].



\bibitem{Witten:1984rs}
E.~Witten,
  {\color{rossoCP3} Cosmic separation of phases},
Phys. Rev. D \textbf{30}, 272-285 (1984)
doi:10.1103/PhysRevD.30.272


\bibitem{Farhi:1984qu}
E.~Farhi and R.~L.~Jaffe,
  {\color{rossoCP3} Strange matter},
Phys. Rev. D \textbf{30}, 2379 (1984)
doi:10.1103/PhysRevD.30.2379


\bibitem{DeRujula:1984axn}
A.~De Rujula and S.~L.~Glashow,
  {\color{rossoCP3} Nuclearites: A novel form of cosmic radiation},
Nature \textbf{312}, 734-737 (1984)
doi:10.1038/312734a0


\bibitem{Alcock:1988re}
C.~Alcock and A.~Olinto,
  {\color{rossoCP3} Exotic phases of hadronic matter and their astrophysical application},
Ann. Rev. Nucl. Part. Sci. \textbf{38}, 161-184 (1988)
doi:10.1146/annurev.ns.38.120188.001113


\bibitem{Chin:1979yb}
S.~A.~Chin and A.~K.~Kerman,
 {\color{rossoCP3} Possible longlived hyperstrange multi-quark droplets},
Phys. Rev. Lett. \textbf{43}, 1292 (1979)
doi:10.1103/PhysRevLett.43.1292

\bibitem{Bai:2018dxf}
Y.~Bai, A.~J.~Long and S.~Lu,
  {\color{rossoCP3} Dark quark nuggets},
Phys. Rev. D \textbf{99}, no.5, 055047 (2019)
doi:10.1103/PhysRevD.99.055047
[arXiv:1810.04360 [hep-ph]].

\bibitem{Jacobs:2014yca}
D.~M.~Jacobs, G.~D.~Starkman and B.~W.~Lynn,
{\color{rossoCP3} Macro dark matter},
Mon. Not. Roy. Astron. Soc. \textbf{450}, no.4, 3418-3430 (2015)
doi:10.1093/mnras/stv774
[arXiv:1410.2236 [astro-ph.CO]].


\bibitem{Sidhu:2018auv}
J.~Singh Sidhu, R.~M.~Abraham, C.~Covault and G.~Starkman,
 {\color{rossoCP3} Macro detection using fluorescence detectors},
JCAP \textbf{02}, 037 (2019)
doi:10.1088/1475-7516/2019/02/037
[arXiv:1808.06978 [astro-ph.HE]].



\bibitem{Bacholle:2020emk}
S.~Bacholle, \textit{et al.},
  {\color{rossoCP3} Mini-EUSO Mission to Study Earth UV Emissions on board the ISS},
Astrophys. J. Suppl. \textbf{253}, no.2, 36 (2021)
doi:10.3847/1538-4365/abd93d
[arXiv:2010.01937 [astro-ph.IM]].


\bibitem{Wiencke:2019vke}
L.~Wiencke and A. Olinto [for the JEM-EUSO and POEMMA Collaborations],
 {\color{rossoCP3} The Extreme Universe Space Observatory on a super-pressure balloon II mission},
PoS \textbf{ICRC2019}, 466 (2020)
doi:10.22323/1.358.0466
[arXiv:1909.12835 [astro-ph.IM]].

\bibitem{Olinto:2020oky}
A.~V.~Olinto \textit{et al.} [POEMMA Collaboration],
 {\color{rossoCP3}  The POEMMA (Probe of Extreme Multi-Messenger Astrophysics) Observatory},
[arXiv:2012.07945 [astro-ph.IM]].


\bibitem{Anchordoqui:2018qom}
L.~A.~Anchordoqui,
 {\color{rossoCP3} Ultra-high-energy cosmic rays},
Phys. Rept. \textbf{801}, 1-93 (2019)
doi:10.1016/j.physrep.2019.01.002
[arXiv:1807.09645 [astro-ph.HE]].

\bibitem{Weinberg:2008zzc}
S.~Weinberg,
{\color{rossoCP3} Cosmology},
(Oxford University Press, 2008)
ISBN:9780198526827

\bibitem{Adams:2014vgr}
J. H. Adams \textit{et al.} [JEM-EUSO Collaboration],
{\color{rossoCP3} JEM-EUSO: Meteor and nuclearite observations},
Exper. Astron. \textbf{40}, no.1, 253-279 (2015)
doi:10.1007/s10686-014-9375-4
  

\bibitem{Paul:2021bhh}
T.~C.~Paul, S.~T.~Reese, L.~A.~Anchordoqui and A.~V.~Olinto,
 {\color{rossoCP3} EUSO-SPB2 sensitivity to macroscopic dark matter},
[arXiv:2104.01152 [hep-ph]].



\bibitem{Sidhu:2019fgg}
J.~S.~Sidhu and G.~Starkman,
 {\color{rossoCP3}  Macroscopic dark matter constraints from bolide camera networks},
Phys. Rev. D \textbf{100}, no.12, 123008 (2019)
doi:10.1103/PhysRevD.100.123008
[arXiv:1908.00557 [astro-ph.CO]].

\end{thebibliography}
\end{document}